\documentclass[12pt]{article}
\usepackage[T1]{fontenc}
\usepackage{lmodern}
\usepackage{microtype}
\usepackage[a4paper,margin=2.2cm]{geometry}
\usepackage{hyperref}
\usepackage[utf8]{inputenc}
\usepackage{graphicx}
\usepackage{amsmath}
\usepackage{cite}%
\usepackage{amsfonts}%
\usepackage{amssymb}
\newcommand \be{\begin{equation}}
\newcommand \ee{\end{equation}}

\begin{document}

\title{Electrostatics of a Finite Conducting Cylinder: Elliptic-Kernel Integral
Equation and Capacitance Asymptotics}
\author{J. Ricardo de Sousa\\Universidade Federal do Amazonas,\\Departamento de F\'{\i}sica, \\3000, Japiim, 69077-000, Manaus-AM, Brazil.}
\date{}
\maketitle

\begin{abstract}
We study the electrostatics of a thin, finite-length conducting cylindrical
shell held at constant potential $V_{0}$. Exploiting axial symmetry, we recast
the problem as a one-dimensional singular integral equation for the axial
surface-charge density, with a kernel written in terms of complete elliptic
integrals. A Chebyshev-weighted collocation scheme that incorporates the
square-root edge singularity yields rapidly convergent charge profiles and
dimensionless capacitances for arbitrary aspect ratios $L/a$, recovering known
long- and short-cylinder limits and providing accurate benchmark values in the
intermediate regime. The method offers a compact, numerically robust reference
formulation for the electrostatics of finite cylindrical conductors.

\end{abstract}

\section{Introduction}

The electrostatics of finite conductors remains a canonical benchmark for both
analytical methods in potential theory and high-accuracy numerical solvers.
Among such geometries, the finite right-circular cylinder, equivalently, an
open conducting tube, occupies a particularly important position, as it
continuously interpolates between two singular limits: the slender-body
regime, characterized by weak axial variation of the surface charge except
near the ends, and the short-cylinder or ring-like regime, in which a
vanishing length at fixed radius leads to behavior dominated by rim effects.

The capacitance and surface-charge distribution of finite-length cylinders
constitute a classic electrostatic problem that admits no closed-form solution
in elementary functions and has therefore been investigated extensively by
numerical and semi-analytical means. Early experimental and theoretical
interest dates back to classical capacitance measurements and bounds
associated with Cavendish and Maxwell\cite{c1,maxwell}. Since then, the
problem has remained a stringent test case, combining axial symmetry with a
nontrivial edge singularity that challenges both analysis and
computation\cite{c2}.

As a result, the open-cylinder geometry has attracted sustained attention in
the applied-electromagnetics and electrostatics communities.
Mid-twentieth-century studies employed boundary-integral and matrix methods,
producing numerical data and practical approximations across wide ranges of
aspect ratio. In this context, the work of Vainshtein and related
contributions by Kapitsa, Fock, and collaborators\cite{c2a,c2b,c2c} are widely
regarded as milestones in the systematic numerical analysis of hollow finite
cylinders\cite{c3}. Subsequent developments included dual integral equation
formulations, often reducible to linear systems via Neumann-series-type
constructions, method-of-moments implementations, and semi-empirical
parameterization for engineering use. Verolino, in particular, formulated the
surface-charge-density problem for a hollow metallic cylinder within the
framework of dual integral equations and provided a detailed assessment of
classical approximations\cite{c4}.

Parallel efforts focused on the capacitance of the open cylinder as a global
observable. Classical engineering-level formulas and numerical comparisons
were reported early on, and later studies refined these approximations and
tabulated high-accuracy values spanning both the tube and ring limits.
Scharstein's Capacitance of a Tube is frequently cited in this context, and
modern reviews often list it alongside earlier analytical and numerical
results\cite{c5}. More recent work has proposed analytic and semi-analytic
representations that combine explicit singular terms with rapidly convergent
expansions, such as Legendre-series constructions, yielding closed-form or
near--closed-form expressions consistent with established
benchmarks\cite{c6,exp1,exp2,exp3,exp4,exp5,exp6,exp7,exp8}.

Despite this substantial literature, two practical gaps remain. First, many
studies emphasize capacitance values or low-parameter fits, whereas
publication-quality, high-resolution surface-charge distributions for a
prescribed constant potential are less commonly reported in a form that is
directly reproducible and numerically stable across broad aspect ratios.
Second, even when charge-density information is available, the integrable
endpoint singularity at the rims can obscure convergence analyses unless the
discretization explicitly incorporates the correct edge behavior.

From a mathematical standpoint, the difficulty lies not in solving Laplace's
equation itself, but in handling the mixed (dual) boundary character
associated with an open surface possessing sharp rims. The surface charge
density on a finite cylindrical shell exhibits a universal square-root
divergence at the edges, and numerical schemes that fail to encode this
behavior explicitly typically converge slowly and may become ill-conditioned.
This feature closely parallels the broader literature on charged wires,
needles, and thin conductors, where carefully designed benchmark computations
were historically required to resolve competing approximations.

In this work, we address these issues for a zero-thickness conducting
cylindrical shell of radius $a$ and finite length $L$, held at a uniform
potential $V_{0}$. Our primary goal is to construct a numerically robust
real-space formulation that resolves the edge singularities and remains well
conditioned over a broad range of aspect ratios. To that end, we first derive
an exact one-dimensional axisymmetric integral equation in which the surface
potential is expressed through a nonlocal operator with a kernel involving
complete elliptic integrals, arising from azimuthal integration of
ring-to-ring interactions. We then solve this singular integral equation by
means of a Chebyshev-weighted collocation scheme that factors out the rim
singularity by construction. This yields rapidly convergent, stable solutions
for both the surface-charge density and the dimensionless capacitance
$\widetilde{C}(\alpha)=C/(2\pi\varepsilon_{0}a)$, with $\alpha=a/L$, and
allows a systematic finite-size scaling analysis to control discretization
errors and extrapolate to the continuum limit. Only after establishing this
real-space solution do we revisit the classical Bessel-based
dual-integral-equation (spectral) representation and show that it generates
exactly the same elliptic-kernel operator, thereby providing an independent
analytical validation and clarifying how distinct mathematical approaches
encode the same physical content\cite{c4}.

The paper is organized as follows. In Section 2 we introduce the physical
model and derive the elliptic-kernel integral equation governing the axial
surface-charge density. In Section 3 we describe the Chebyshev-weighted
numerical method. Section 4 discusses numerical results for the surface-charge
density and capacitance across a wide range of aspect ratios, including the
extrapolation procedure used to obtain continuum-limit results and asymptotic
regimes. Section 5 presents the equivalent spectral formulation based on dual
integral equations and establishes explicitly the correspondence between the
two approaches. Finally, Section 6 summarizes the main conclusions and
outlines possible extensions.

\section{Physical Model and Integral Equation}

We consider a thin (zero-thickness) conducting cylindrical shell aligned with
the z$-$axis, of radius $a$ and finite length $L$, occupying the interval
$-L/2<z<L/2$. The conductor is held at a uniform electrostatic potential
$V_{0}$, as sketched in Fig. 1. In the absence of external fields, the
configuration is axially symmetric; therefore the induced surface-charge
density is independent of the azimuthal angle $\varphi$ and depends only on
the axial coordinate, i. e., $\sigma(z)$, for $\rho=a$ and $z\in\left[
-L/2,L/2\right]  $.

\begin{figure}%
[p]
  \centering
  \includegraphics[width=0.78\linewidth]{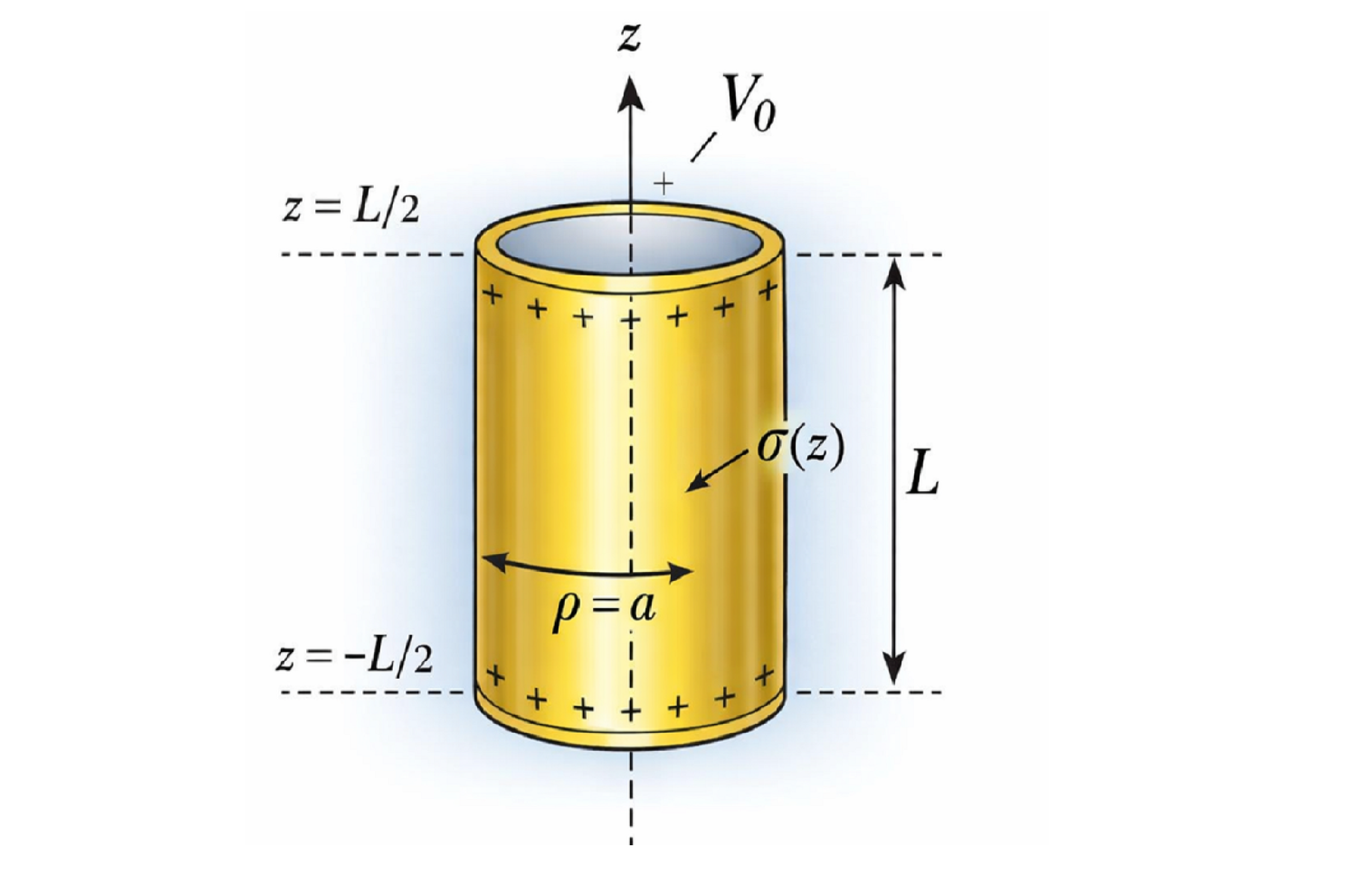}%

  \caption{Schematic representation of a thin conducting cylindrical shell of radius $a$ and finite length $L$, aligned with the $z$-axis and held at a uniform electrostatic potential $V_0$. The cylinder occupies the region $-L/2<z<L/2$ and has zero thickness. Owing to axial symmetry and the absence of external fields, the induced surface-charge density is independent of the azimuthal angle and depends only on the axial coordinate, $\sigma=\sigma(z)$, on the surface $\rho=a$.}%

  \label{fig:geometry_thin}
\end{figure}

The objective of this work is to determine the surface-charge density
$\sigma(z)$ induced on a finite conducting cylindrical shell held at a
constant electrostatic potential. Although this is a classical problem in
electrostatics, it still warrants further analysis due to its nontrivial
mathematical structure and the subtle role played by edge effects.

We employ two complementary analytical approaches. In the first approach, a
more direct formulation is adopted, in which the potential is expressed
through real-space integral representations involving \textit{complete
elliptic integrals}. In the second, a \textit{spectral formulation} is
developed by expanding the electrostatic potential in terms of Bessel
functions, which naturally leads to a system of dual integral equations for
the unknown charge distribution.

We explicitly demonstrate that these two formulations are mathematically
equivalent, providing different but consistent perspectives on the same
physical problem. This equivalence not only clarifies the underlying structure
of the solution but also highlights the connections between spectral methods,
nonlocal integral operators, and classical kernel representations in electrostatics.

In cylindrical coordinates, the electrostatic potential at an arbitrary field
point ($\rho,\varphi,z$) due to a surface-charge density distributed on the
lateral surface $\rho^{\prime}=a$ is%
\begin{equation}
\Phi(\rho,z)=k_{e}\int\limits_{-L/2}^{L/2}\int\limits_{0}^{2\pi}%
\frac{\sigma(z^{\prime})adz^{\prime}d\varphi^{\prime}}{\sqrt{\rho^{2}%
+a^{2}+\left(  z-z^{\prime}\right)  ^{2}-2a\rho\cos(\varphi-\varphi^{\prime}%
)}}, \label{c1}%
\end{equation}
where $k_{e}=1/(4\pi\varepsilon_{0})$ is the electrostatic constant. Since
$\sigma(z^{\prime})$ is independent of $\varphi^{\prime}$, the azimuthal
integral can be performed analytically in terms of the complete elliptic
integral of the first kind $K(m)$, yielding%
\begin{equation}
\Phi(\rho,z)=4k_{e}a\int\limits_{-L/2}^{L/2}\sigma(z^{\prime}%
)\frac{K[m(z^{\prime})]}{\sqrt{\left(  \rho+a\right)  ^{2}+\left(
z-z^{\prime}\right)  ^{2}}}dz^{\prime}, \label{c2}%
\end{equation}
with%
\begin{equation}
K[m(z^{\prime})]=\int\limits_{0}^{\pi/2}\frac{d\theta}{\sqrt{1-m(z^{\prime
})\sin^{2}(\theta)}}, \label{c3}%
\end{equation}
and elliptic parameter%
\begin{equation}
m(z^{\prime})=\frac{4a\rho}{\left(  \rho+a\right)  ^{2}+\left(  z-z^{\prime
}\right)  ^{2}}. \label{c4}%
\end{equation}

Imposing the Dirichlet boundary condition on the conductor, $\Phi(a,z)=V_{0}
$, Eq. (\ref{c2}) reduces to the integral equation%
\begin{equation}
\int\limits_{-L/2}^{L/2}\sigma(z^{\prime})\mathcal{G}(z-z^{\prime})dz^{\prime
}=\frac{\pi\varepsilon_{0}}{a}V_{0},\left|  z\right|  <L/2, \label{c5}%
\end{equation}
where the kernel is%
\begin{equation}
\mathcal{G}(q)=\frac{K\left(  \frac{4a^{2}}{4a^{2}+q^{2}}\right)  }%
{\sqrt{4a^{2}+q^{2}}}. \label{c6}%
\end{equation}
Equation (\ref{c5}) is precisely the mathematical core of the problem:
determine $\sigma(z)$ such that the convolution with $\mathcal{G}(q)$ is
constant over $\left|  z\right|  <L/2$. The kernel has a weak (logarithmic)
singularity as $q\rightarrow0$%

\begin{equation}
\mathcal{G}(q\rightarrow0)\simeq\frac{1}{2a}\ln\left(  \frac{8a}{\left|
q\right|  }\right)  +\mathcal{O}(1), \label{c7}%
\end{equation}
which implies an edge divergence of the surface charge. Near the cylinder ends
$z\rightarrow\pm L/2$, one expects the generic square-root blow-up\cite{c2}%

\begin{equation}
\sigma(z)\simeq\frac{\sigma_{0}}{\sqrt{\left(  \frac{L}{2a}\right)
^{2}-\left(  \frac{z}{a}\right)  ^{2}}}, \label{c8}%
\end{equation}
where $\sigma_{0}$ is a constant with the dimensions of surface-charge density.

\section{Chebyshev-Weighted Numerical Method}

Because of this integrable but strong edge divergence, a naive discretization
of Eq. (\ref{c5}) converges slowly and may become numerically ill-conditioned.
To stabilize the computation and enforce the correct endpoint behavior by
construction, we introduce a Chebyshev-weighted parametrization. Using the
dimensionless coordinate $x=(2/L)z\in\left[  -1,1\right]  $, we rewrite the
surface-charge density as%

\begin{equation}
\sigma(z)=\frac{1}{\sqrt{1-x^{2}}}p(x), \label{c9}%
\end{equation}
where the weight $\left(  1-x^{2}\right)  ^{-1/2}$ captures the expected
square-root divergence at $x=\pm1$, while $p(x)$ remains smooth on $\left[
-1,1\right]  $. When convenient, we represent $p(x)$ by a truncated
\textit{Chebyshev expansion in polynomials} of the first kind $T_{n}%
(x)$\cite{tn1,tn2,tn3}
\begin{equation}
p(x)=\sum\limits_{n=0}^{N}c_{n}T_{n}(x), \label{c10}%
\end{equation}
where $N$ is the \textbf{spectral truncation order} and the coefficients
$\{c_{n}\}$ are obtained numerically. The resulting integral equation is
solved by a collocation method combined with Gauss--Chebyshev quadrature,
following standard spectral discretization strategies for weakly singular
kernels\cite{tn4,tn5,tn6}.

With this transformation, Eq. (\ref{c5}) becomes the following integral
equation over $x$:%
\begin{equation}
\int\limits_{-1}^{1}\frac{p(x^{\prime})}{\sqrt{1-x^{\prime2}}}\frac{K\left[
\frac{16\alpha^{2}}{16\alpha^{2}+\left(  x-x^{\prime}\right)  ^{2}}\right]
}{\sqrt{16\alpha^{2}+\left(  x-x^{\prime}\right)  ^{2}}}=\frac{\pi
\varepsilon_{0}}{a}V_{0}, \label{c11}%
\end{equation}
where $\alpha=a/L$ is the dimensionless geometric parameter.

We solve Eq. (\ref{c11}) by a Nystr\"{o}m/collocation discretization on
$\left[  -1,1\right]  $ combined with Gauss--Chebyshev quadrature tailored to
the Chebyshev weight. Specifically:

1. \textbf{Collocation grid}. We partition the interval $x\in\left[
-1,1\right]  $ into $N_{p}$ panels and enforce Eq. (\ref{c11}) at a set of
collocation points $\left\{  x_{j},j=1,2,..N_{c}\right\}  $ (either the
Chebyshev--Lobatto nodes or the panel midpoints mapped to $\left[
-1,1\right]  $).

2. \textbf{Quadrature}. On each panel, the integral over $x^{\prime}$ is
evaluated using an $n_{g}-$point Gauss--Chebyshev rule, which is naturally
adapted to the weight $\left(  1-x^{2}\right)  ^{-1/2}$. Here $n_{g}$ controls
the within-panel quadrature order.

3. \textbf{Linear system}. The discretization yields the dense linear system%
\begin{equation}
\sum\limits_{j=1}^{N_{c}}A_{ij}p(x_{j})=B_{i}, \label{c11a}%
\end{equation}
where $N_{c}$ is the number of collocation conditions (equivalently, the
number of unknown nodal values $p(x_{j})$). In our implementation we collocate
at panel midpoints, so that $N_{c}=N_{p}$. The matrix entries $A_{ij}$ are
assembled from the elliptic-integral kernel evaluated at these nodes together
with the corresponding Gauss--Chebyshev quadrature weights, while $B_{i}$
represents the prescribed constant potential. We solve for $\left\{
p(x_{j})\right\}  $ using standard dense linear-algebra routines.

4. \textbf{Dimensionless normalization}. We report the dimensionless density
$\widetilde{\sigma}(z)=a\sigma(z)/(\varepsilon_{0}V_{0})$, which removes the
trivial dependence on $V_{0}$ and $\varepsilon_{0}$ and isolates the geometric
dependence through $\alpha=a/L$.

\section{Numerical Results}

In order to illustrate the computations presented below, we discretize the
integral equation using $N_{p}=220$ panels in $x$, with $n_{g}=16$
Gauss--Chebyshev points per panel. Figure 2 displays $\widetilde{\sigma}(z)$
as a function of the normalized coordinatef $z/L$ for aspect ratios
$\alpha=1/3,1.0,2.0,$ and $5.0$. In all cases, the surface-charge density
increases monotonically as $z\rightarrow\pm L/2$, in agreement with the edge
divergence predicted by Eq. (\ref{c8}). As $\alpha$ increases, that is, as the
cylinder becomes shorter, the two ends approach each other in the
dimensionless coordinate $z/L$, and edge effects penetrate more deeply into
the interior region, leading to a higher surface-charge density throughout the cylinder.

\begin{figure}%
[p]
  \centering
  \includegraphics[width=0.82\linewidth]{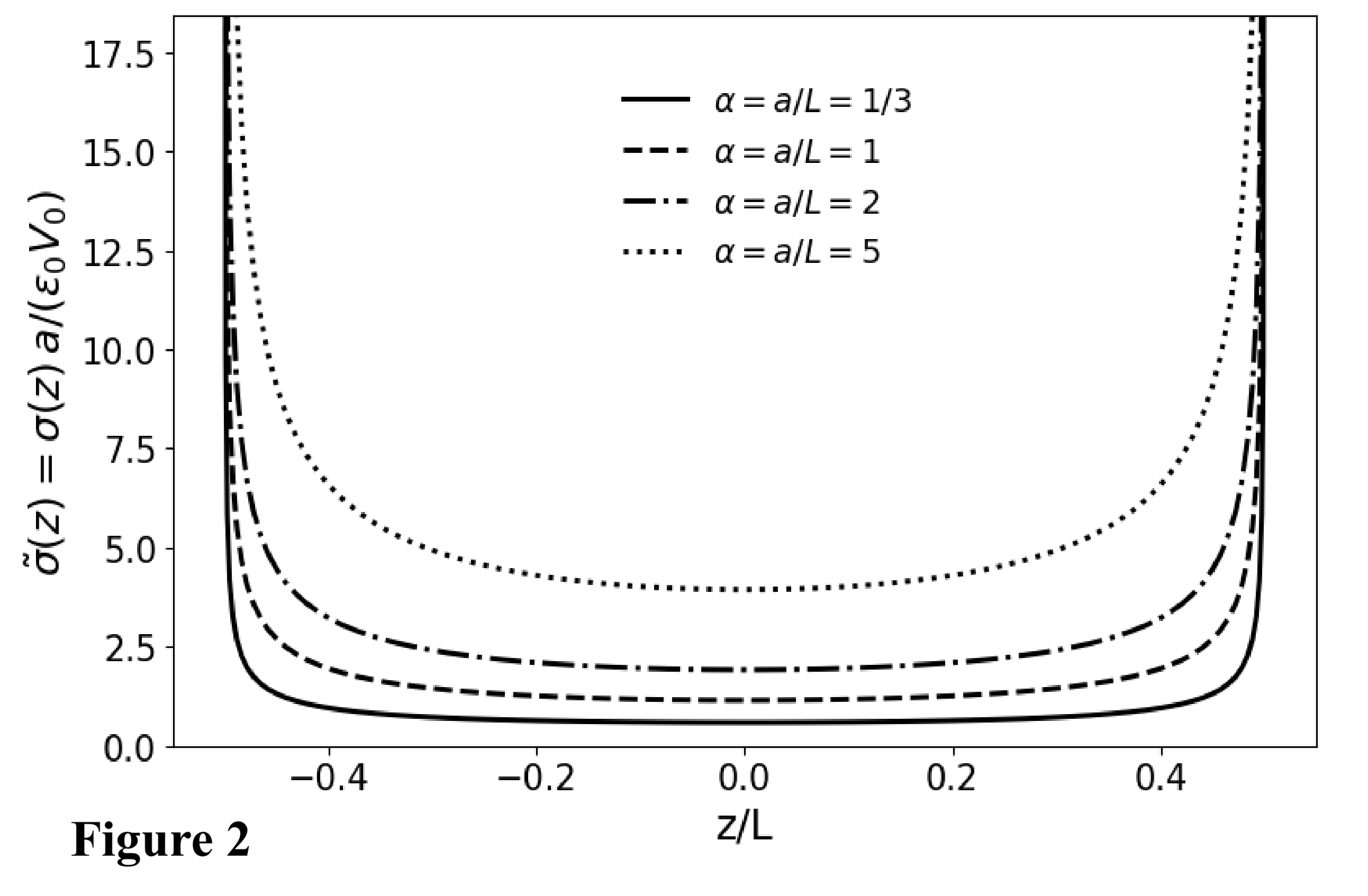}%

  \caption{Dimensionless surface-charge density on the lateral wall of an open conducting cylinder held at a uniform potential $V_0$. The curves show $\tilde{\sigma}%
(z)=a\,\sigma(z)/(\varepsilon_0 V_0)$ as a function of the normalized axial coordinate $z/L$, for four aspect ratios $\alpha=1/3,\,1,\,2,$ and $5$ (different line styles). The vertical scale is restricted to emphasize the interior region; the rapid rise near $z/L=\pm 1/2$ reflects the edge (rim) charge crowding at the open ends.}%

  \label{fig:sigma_profiles_thin}
\end{figure}

At the cylinder center $z=0$, we obtain%
\begin{equation}
\left\{
\begin{array}
[c]{c}%
\widetilde{\sigma}(\alpha=1/3)\simeq0.5844\\
\widetilde{\sigma}(\alpha=1)\simeq1.1466\\
\widetilde{\sigma}(\alpha=2)\simeq1.9160\\
\widetilde{\sigma}(\alpha=5)\simeq3.9371
\end{array}
\right.  . \label{c12}%
\end{equation}

Beyond the qualitative trends summarized above, the main contribution of the
present note is methodological: Eq. (\ref{c5}) provides an exact
boundary-integral formulation with an elliptic-integral kernel, but its weak
\textit{logarithmic singularity} makes direct discretizations poorly
conditioned. By extracting the endpoint behavior analytically through the
weighted representation $\sigma(z)=\left(  1-x^{2}\right)  ^{-1/2}p(x)$, the
remaining unknown $p(x)$ becomes smooth and can be determined accurately by
collocation with Gauss--Chebyshev quadrature. This produces a stable, rapidly
convergent procedure that resolves the edge-dominated regime without ad hoc
smoothing or endpoint fitting, and yields a reproducible reference solution
for $\sigma(z)$ at fixed potential.

\subsection{Finite size-scaling}

To estimate the continuum-limit value of the center density, we compute
$\widetilde{\sigma}(0)$ for a sequence of increasingly refined Chebyshev-panel
discretizations, while keeping the quadrature order within each panel fixed.
Specifically, for each $N_{c}$ we solve the discretized integral equation for
the auxiliary unknown $p(x)$ using $n_{g}=16$ Gauss--Chebyshev points per
panel to evaluate the panel integrals. The center value $\widetilde{\sigma
}(0)$ is then obtained by interpolation from the collocation nodes.

In the asymptotic regime we fit the data to the leading finite$-N_{c}$ form%

\begin{equation}
\widetilde{\sigma}(0,N_{c})=\widetilde{\sigma}_{\infty}(0)+\frac{A}{N_{c}%
}+\mathcal{O}(N_{c}^{-2}),\label{c11b}%
\end{equation}
which is equivalent to a linear regression of $\widetilde{\sigma}(0)$ versus
$1/N_{c}$. Restricting the fit to sufficiently large $N_{c}$ (here $N_{c}%
\geq220$) suppresses pre-asymptotic curvature and yields a stable intercept
$\widetilde{\sigma}_{\infty}(0)$, which we take as our best estimate of the
$N_{c}\rightarrow\infty$ limit.

\begin{figure}%
[p]
  \centering
  \includegraphics[width=0.80\linewidth]{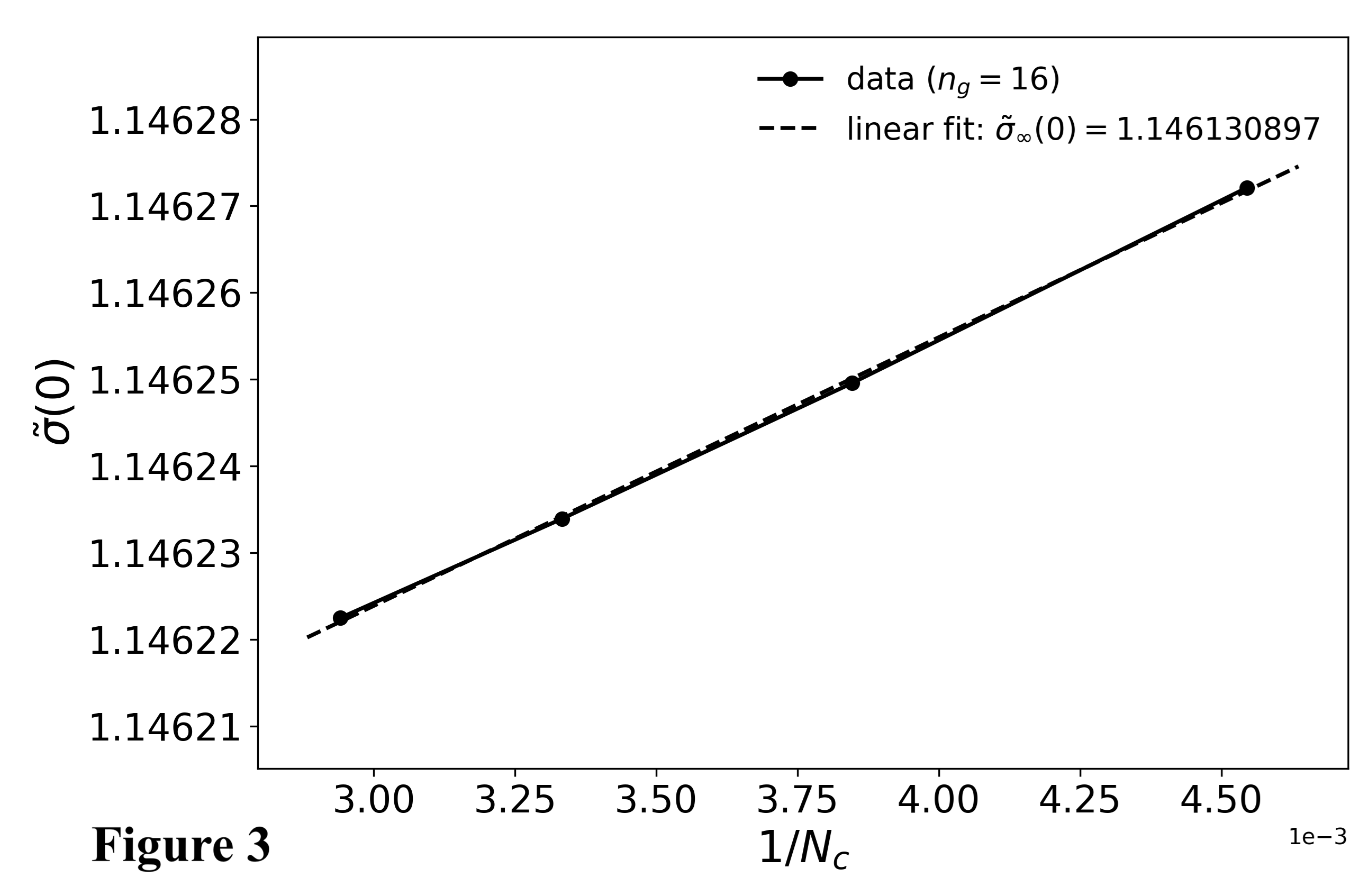}%

  \caption{Finite-size scaling of the numerically computed cylinder-center surface-charge density. The dimensionless center value $\tilde{\sigma}%
(0)$ for $\alpha=1$ is plotted versus $1/N_c$, where $N_c$ is the number of Chebyshev panels used in the axial discretization of the integral equation, with within-panel quadrature fixed at $n_g=16$ Gauss--Chebyshev points. In the asymptotic regime ($N_c\ge 220$) the data fall on an approximately straight line, consistent with a leading discretization error $\tilde{\sigma}%
(0,N_c)=\tilde{\sigma}_\infty(0)+A/N_c+\mathcal{O}(N_c^{-2}%
)$. The dashed line is the corresponding linear least-squares fit, whose intercept provides the $N_c\to\infty$ extrapolation $\tilde{\sigma}%
_\infty(0)$.}
  \label{fig:scaling_sigma0_thin}
\end{figure}

Figure 3 displays a finite-size scaling analysis of the numerical collocation
scheme. We plot the cylinder-center density $\widetilde{\sigma}(0)$ as a
function of $1/N_{c}$. A least-squares extrapolation based on the sequence of
discretizations $N_{c}=\{220,260,300,340\}$ yields the converged value
$\widetilde{\sigma}_{\infty}(0)\approx1.1461309$ for $\alpha=1$. The same
extrapolation can be carried out for any aspect ratio $\alpha=a/L$. In
practice, we find that the $N_{c}\rightarrow\infty$ intercepts differ only
weakly from the direct $N_{c}=220$ results: for the range of $\alpha$
considered in this work, the correction $\Delta\widetilde{\sigma}(0)=\left|
\widetilde{\sigma}(0,N_{c}=220)-\widetilde{\sigma}_{\infty}(0)\right|  $ is
small compared with the overall variation of $\widetilde{\sigma}(0)$ with
$\alpha$. Thus, while Fig. 3 illustrates the procedure explicitly for
$\alpha=1$, the values reported throughout the paper at $N_{c}=220$ already
provide an excellent approximation to the extrapolated continuum-limit
densities, and the finite$-N_{c}$ extrapolation mainly serves as a
quantitative validation of numerical convergence.

These results are significant because the surface density directly sets the
local normal field, $E_{n}(z)=\sigma(z)/\varepsilon_{0}$, so the computed
profiles quantify where the maximum fields occur and how strongly they
concentrate near the ends as geometry is varied. In practical electrode
design, such end concentrations often determine breakdown and emission
thresholds; therefore, an accurate benchmark for $\sigma(z)$ provides a
controlled way to estimate peak-field enhancement in finite cylindrical
electrodes. Moreover, the present framework, exact kernel plus endpoint-aware
discretization, can be extended with minimal changes to segmented electrodes,
nonuniform boundary potentials, or coupled conductor configurations, making it
a useful computational template rather than a one-off calculation.

\subsection{Capacitance}

An additional quantity of direct interest that follows directly from the
numerical solution for the surface-charge density $\sigma(z)$ is the
self-capacitance of the finite cylindrical conductor. For an isolated
conductor held at a uniform potential $V_{0}$, the capacitance provides a
compact global measure of the electrostatic response and is particularly
useful for comparison with classical results and asymptotic approximations.

We define the geometry-dependent capacitance coefficient $C_{11}\equiv
C(\alpha)$ through the total charge on the lateral surface,%

\begin{equation}
C(\alpha)=\frac{1}{V_{0}}\int\limits_{-L/2}^{L/2}\sigma(z^{\prime})2\pi
adz^{\prime}, \label{c13}%
\end{equation}
so that, by construction, the total charge satisfies $Q=C(\alpha)V_{0}$. The
corresponding electrostatic self-energy is then given by%

\begin{equation}
U(\alpha)=\frac{1}{2}C(\alpha)V_{0}^{2}, \label{c13a}%
\end{equation}
which provides an energetically meaningful characterization of the system and
offers an alternative global diagnostic of the numerical solution.

Because the capacitance integrates the full surface-charge distribution, it is
sensitive to both the bulk behavior and the rim singularities, and thus serves
as a stringent test of numerical accuracy. Moreover, its dependence on the
aspect ratio $\alpha=a/L$ allows for direct assessment of the crossover
between the long-cylinder and short-cylinder regimes, facilitating comparison
with known asymptotic results and previously reported benchmark values.

\subsubsection{Asymptotic Regimes}

Before proceeding to the full numerical solution of Eq. (\ref{c13}), it is
instructive to examine the physically relevant asymptotic limits of the hollow
cylindrical capacitor. These limiting regimes provide valuable insight into
the structure of the surface-charge distribution, clarify the origin and role
of edge singularities, and serve as analytical benchmarks against which the
numerical results can be assessed. In particular, the asymptotic analysis
elucidates how the problem interpolates between the long-cylinder
(slender-body) limit and the short-cylinder (ring-like) limit, thereby
highlighting the distinct physical mechanisms that dominate the electrostatics
in each case.

\paragraph{Long cylinder ($\alpha<<1$)}

In the slender-cylinder limit, $\alpha<<1$ ($L>>a$), the capacitance shows the
familiar logarithmic dependence associated with three-dimensional conductors
with a small transverse length scale. A convenient derivation follows from a
slender-body approximation: away from the rims the charge varies slowly, so
the cylinder may be modeled by a nearly uniform line charge of linear density
$\lambda\simeq Q/L$. The resulting potential on the surface, evaluated at a
representative point (e.g. near the midplane), has the asymptotic form%
\begin{equation}
V_{0}\simeq\frac{\lambda}{2\pi\varepsilon_{0}}\left[  \ln\left(  \frac{2L}%
{a}\right)  -1\right]  , \label{c12a}%
\end{equation}
where the dominant $\ln(2L/a)$ term is universal and the additive constant
accounts for end corrections in the three-dimensional geometry
(\textit{Maxwell's classical form})\cite{maxwell}. Solving for $C=Q/V_{0}$ yields%

\begin{equation}
\widetilde{C}(\alpha)=\frac{C(\alpha)}{2\pi\varepsilon_{0}a}\simeq
\frac{\left(  1/\alpha\right)  }{\ln\left(  2/\alpha\right)  -1}\text{,
}\alpha<<1. \label{c14}%
\end{equation}
This expression highlights the ``almost linear'' growth of capacitance with
the length $L$ (since $1/\alpha=L/a$), corrected by the slowly varying
logarithmic factor $\ln\left(  2/\alpha\right)  $. Although $\ln\left(
2/\alpha\right)  \rightarrow\infty$ as $\alpha\rightarrow0$, it does so only
slowly, whereas $1/\alpha$ grows unbounded; the net effect is a divergence of
$\widetilde{C}(\alpha)$ as $\alpha\rightarrow0$, weaker than purely linear
scaling but still without bound in the idealized limit.

Below we report representative numerical values of the dimensionless
capacitance for $N_{c}=220$. The results clearly reveal the singular behavior
of the capacitance as $\alpha\rightarrow0$, corresponding to the thin,
long-cylinder limit. Specifically, we find%

\begin{equation}
\left\{
\begin{array}
[c]{c}%
\widetilde{C}_{\infty}(\alpha=0.1)\simeq4.96\\
\widetilde{C}_{\infty}(\alpha=0.05)\simeq7.58\\
\widetilde{C}_{\infty}(\alpha=0.02)\simeq14.20\\
\widetilde{C}_{\infty}(\alpha=0.01)\simeq23.73
\end{array}
\right.  . \label{c13b}%
\end{equation}
These values demonstrate the rapid growth of the capacitance as the aspect
ratio decreases, reflecting the increasing dominance of end effects and the
associated divergence of the surface-charge density in this regime.

Many classic tabulations report the electrostatic capacity of finite cylinders
in Gaussian units, where capacitance has dimensions of length and is commonly
quoted as $C/a$ versus $x=L/a$. In particular, Table 1 of Ref.\cite{exp3}
provides high-accuracy benchmark values for the hollow (open) cylinder in the
column $C^{(H)}/a$. To compare these data with our SI-based normalization, we
define the dimensionless capacitance $\widetilde{C}(\alpha)=C_{\text{SI}%
}/(2\pi\varepsilon_{0}a)$, with $\alpha=a/L=1/x$, and use the conversion
$C_{SI}=4\pi\varepsilon_{0}C_{G}$. Since Ref.\cite{exp3} reports
$C_{G}/a=C^{(H)}/a$, the mapping becomes simply $\widetilde{C}(\alpha
)=2C_{G}/a$. Table 1 in the present paper summarizes this conversion and
compares the resulting benchmark values with our numerical results obtained
from the surface-charge solution and total-charge integration, Eq.
(\ref{c13}). For representative slender cylinders ( $x=10,20,50,100$, i. e.,
$\alpha=0.1,0.05,0.02,0.01$), the agreement remains at the sub-percent level,
thereby providing an external validation of both the computed charge density
on the lateral surface and the capacitance extraction procedure.\newline 

\textbf{Table 1}: Benchmark of the dimensionless capacitance for an open
hollow conducting cylinder (lateral surface only). Reference values are taken
from Table 1 of Ref.\cite{exp3} (column $C^{(H)}/a$) as a function $x=L/a$; we
use $\alpha=a/L=1/x$ to match our notation. To convert the Ref.\cite{exp3}
data to SI, $C_{SI}=4\pi\varepsilon_{0}a(C^{(H)}/a)$; equivalently, in our
normalization $\widetilde{C}(\alpha)=C/(2\pi\varepsilon_{0}a)=2(C^{(H)}/a)$ .
The relative error reported in the last column is $\Delta=\frac{\left|
\widetilde{C}_{\text{\cite{exp3}}}-\widetilde{C}_{\text{this work}}\right|
}{\widetilde{C}_{\text{\cite{exp3}}}} $. The agreement is at the sub-percent
level over the tested aspect ratios.

\begin{center}%
\begin{tabular}
[c]{llllll}\hline\hline
$x=L/a$ & $\alpha=a/L$ & $C^{(H)}/a$\cite{exp3} & $\widetilde{C}%
(\alpha)=(2/a)C^{(H)}$\cite{exp3} & This work & $\Delta(\%)$\\\hline\hline
$10$ & $0.1$ & $2.479711$ & $4.959422$ & $4.960616$ & $0.024$\\
$20$ & $0.05$ & $3.788663$ & $7.577326$ & $7.580178$ & $0.038$\\
$50$ & $0.02$ & $7.092673$ & $14.185346$ & $14.195725$ & $0.073$\\
$100$ & $0.01$ & $11.854900$ & $23.709800$ & $23.739575$ & $0.126$%
\\\hline\hline
\end{tabular}
\end{center}

The residual sub-percent discrepancies in Table 1 are expected and have two
closely related origins. First, the benchmark values in Ref.\cite{exp3} were
obtained using a different high-precision numerical formulation (Love-type
integral-equation methods and related schemes), whereas we solve an
elliptic-kernel boundary integral equation using a Chebyshev-weighted
parametrization combined with Nystr\"{o}m/collocation and panel quadrature.
Since the two approaches rely on different discretizations, their truncation
and quadrature errors are not identical. In our implementation the dominant
numerical uncertainty is controlled by the finite-resolution parameters
($N_{c},n_{g}$) and by the accurate handling of the integrable rim
singularity, which strongly influences convergence of the total-charge
integral used to extract $\widetilde{C}(\alpha)$, Eq. (\ref{c13}).

Second, the magnitude of the discrepancies is consistent with the
finite$-N_{c}$ effects quantified independently in Sec. 4.1. As illustrated by
the finite-size scaling analysis of the center density in Fig. 3 and the
leading correction form Eq. (\ref{c11b}), the $N_{c}\rightarrow\infty$
extrapolation differs only weakly from the direct $N_{c}=220$ results. Because
the capacitance is obtained from the same numerical charge density via a
global integration, it inherits a discretization error of comparable order
(set primarily by $N_{c}$ at fixed $n_{g}$). Therefore, the observed offsets
in Table 1 are naturally attributed to residual $\mathcal{O}(1/N_{c}) $ (and
subleading $\mathcal{O}(N_{c}^{-2})$) discretization errors in the numerical
evaluation of the integral operator and in the rim-resolved charge
integration, rather than to any systematic bias. Increasing $N_{c}$ (or
performing an explicit $N_{c}\rightarrow\infty$ extrapolation for
$\widetilde{C}(\alpha)$) would reduce the remaining differences further.

\paragraph{Short cylinder ($\alpha>>1$)}

In the extreme short-cylinder limit$\ L\rightarrow0$ (i. e., $\alpha
=a/L\rightarrow\infty$), the lateral surface $\rho=a$, $\left|  z\right|
<L/2$ collapses geometrically to a ring-like object. Start from the
axisymmetric Coulomb representation (\ref{c1}) of the surface potential on the
lateral wall ($\rho=a$) in terms of the longitudinal density $\sigma(z) $:%
\begin{equation}
\Phi(a,z)=V_{0}=k_{e}\int\limits_{-L/2}^{L/2}a\sigma(z^{\prime})dz^{\prime
}\int\limits_{0}^{2\pi}\frac{d\varphi^{\prime}}{\sqrt{\left(  z-z^{\prime
}\right)  ^{2}+2a^{2}\sin^{2}(\varphi^{\prime})}}. \label{c13c}%
\end{equation}
The azimuthal integral is expressible through a complete elliptic integral
and, for $\left|  z-z^{\prime}\right|  <<a$ (which holds uniformly when
$L<<a$) its standard near-singular expansion reduces the kernel to a
logarithmic interaction along $z$:%
\begin{equation}
\int\limits_{0}^{2\pi}\frac{d\varphi^{\prime}}{\sqrt{\left(  z-z^{\prime
}\right)  ^{2}+2a^{2}\sin^{2}(\varphi^{\prime})}}\simeq\frac{2}{a}\ln\left(
\frac{8a}{\left|  z-z^{\prime}\right|  }\right)  +\mathcal{O}\left[
\frac{\left(  z-z^{\prime}\right)  ^{2}}{a^{3}}\ln\left(  \frac{8a}{\left|
z-z^{\prime}\right|  }\right)  \right]  . \label{c13d}%
\end{equation}

Hence, to leading order, the electrostatic potential satisfies%
\begin{equation}
V_{0}=2k_{e}\int\limits_{-L/2}^{L/2}\sigma(z^{\prime})\ln\left(
\frac{8a}{\left|  z-z^{\prime}\right|  }\right)  dz^{\prime}+\text{subleading
}\mathcal{O}\left[  \left(  L/a\right)  ^{2}\ln\left(  a/L\right)  \right]  .
\label{c13e}%
\end{equation}

Introducing the dimensionless variable $x=2z/L\in\lbrack-1,1]$ and motivated
by the expected divergence of the surface-charge density at the cylinder ends,
we adopt the leading-order \textit{ansatz}%
\begin{equation}
\sigma(z)\simeq\frac{A}{\sqrt{1-x^{2}}}. \label{c13f}%
\end{equation}
Using the classical identity for the logarithmic potential weighted by the
Chebyshev measure\cite{tab},%

\begin{equation}
\int\limits_{-1}^{1}\frac{\ln\left|  x-x^{\prime}\right|  }{\sqrt{1-x^{2}}%
}dx^{\prime}=-\pi\ln2\text{, }\left|  x\right|  <1. \label{c13g}%
\end{equation}
and noting that $\left|  z-z^{\prime}\right|  =(L/2)\left|  x-x^{\prime
}\right|  $, the integral equation reduces, at leading order and independently
of $x$, to%
\begin{equation}
V_{0}=\frac{LA}{4\varepsilon_{0}}\ln\left(  \frac{32a}{L}\right)  .
\label{c13h}%
\end{equation}
Solving for the amplitude $A$, we obtain%
\begin{equation}
A=\frac{4\varepsilon_{0}V_{0}}{L\ln(32a/L)}. \label{c13i}%
\end{equation}

Substituting Eq. (\ref{c13i}) into the ansatz (\ref{c13f}), we find the
dimensionless surface-charge density
\begin{equation}
\widetilde{\sigma}(z)=\frac{a}{\varepsilon_{0}V_{0}}\sigma(z)\simeq
\frac{4\alpha}{\ln(32\alpha)}\frac{1}{\sqrt{1-x^{2}}}\text{, }x=\frac{2z}%
{L},\alpha=\frac{a}{L}>>1. \label{c13j}%
\end{equation}

Using the asymptotic expression (\ref{c13j}), the total induced charge follows
as%
\begin{equation}
Q=\int\limits_{-L/2}^{L/2}2\pi a\sigma(z^{\prime})dz^{\prime}=2\pi
a\frac{L}{2}A\int\limits_{-1}^{1}\frac{dx}{\sqrt{1-x^{2}}}=\pi^{2}aLA.
\label{c13l}%
\end{equation}
Consequently, the capacitance is given by%
\begin{equation}
C=\frac{Q}{V_{0}}\simeq\frac{4\pi^{2}\varepsilon_{0}a}{\ln(32\alpha)}\text{
(}L<<a\text{),} \label{c13m}%
\end{equation}
and, using the adimensional definition $\widetilde{C}(\alpha)=C/(2\pi
\varepsilon_{0}a)$, we finally obtain%
\begin{equation}
\widetilde{C}(\alpha)\simeq\frac{2\pi}{\ln(32\alpha)}\text{ (}\alpha
>>1\text{).} \label{c13n}%
\end{equation}
This asymptotic result was first obtained by Lebedev and Skal'skaya\cite{exp1}
using the method of dual integral equations.

This \textit{logarithmic law} is consistent with the standard
short-tube$\rightarrow$ ring approximation reported in the capacitance
literature for a hollow cylinder when the length is much smaller than the
radius. In particular, several sources state a leading behavior of the form%
\begin{equation}
C\propto\frac{a}{\ln\left(  \text{const}\times L/a\right)  }, \label{c13o}%
\end{equation}
often written with a half-length notation $\ell=L/2$, leading to $\ln
(16a/\ell)=\ln(32a/L)$\cite{exp1}. High-precision computations for hollow
cylinders also explicitly probe very small $L/a$ and are consistent with this
slow $1/\ln(a/L)$ decay in the short-length
regime\cite{exp2,exp3,exp4,exp5,exp6,exp7}.

A related but conceptually distinct approach to the evaluation of
electrostatic energy and capacitance was recently proposed by Arun, et
al.\cite{exp8}, who introduced an algebraic--topological method for computing
stored electrostatic energy and three-dimensional Maxwellian capacitance in
complex conductor configurations. Their framework emphasizes global
topological invariants and network-based representations of the electrostatic
field, offering an alternative to traditional boundary-integral or spectral
formulations. In contrast, the present work focuses on a direct continuum
description based on integral equations with physically transparent kernels,
allowing for an explicit characterization of the surface-charge density and
its singular behavior. Together, these complementary approaches highlight the
diversity of available theoretical tools for capacitance evaluation and
underscore the relevance of geometry-dependent effects in finite electrostatic systems.

However, for a zero-thickness conducting shell this limit is singular: the
problem ceases to be a strictly two-dimensional surface-conductor
boundary-value problem and effectively approaches a one-dimensional support,
for which the notion of a smooth surface-charge density and a unique
capacitance requires an additional transverse length scale. In practice, a
physical ``ring'' is regularized by a finite wire radius or wall thickness
(e.g., a \textit{thin torus}), and the capacitance becomes finite but depends
logarithmically on that transverse scale. Accordingly, while the ring-like
interpretation is geometrically correct, the $\alpha\rightarrow\infty$ limit
of the present idealized shell should be regarded as a formal asymptote rather
than a universal, thickness-independent prediction; our numerical results are
therefore intended for finite $\alpha$, where the shell model is well posed.
Hence, we do not claim a universal $\alpha\rightarrow\infty$ capacitance for
the zero-thickness model; the asymptote is formal.

For the particular case of a toroidal conductor, with central radius $a$ and a
circular section of radius $R$, the dimensionless capacity is given by
Refs.\cite{t1,t2,t3} as%
\begin{equation}
\widetilde{C}=\frac{8\pi}{a}\sqrt{a^{2}-R^{2}}\left[  \frac{1}{2}%
\frac{Q_{-1/2}(a/R)}{P_{-1/2}(a/R)}+\sum\limits_{n=1}^{\infty}\frac{Q_{n-1/2}%
(a/R)}{P_{n-1/2}(a/R)}\right]  , \label{c13p}%
\end{equation}
where $Q_{s}$ and $P_{s}$ denote the Legendre functions of the first and
second kind, respectively. We have adapted the original expression for the
capacitance, given in Gaussian units, to the SI system. In our notation this
corresponds to the transformation $\widetilde{C}(\alpha)=2C_{G}/a$, as
discussed above.

The asymptotic expansion of (\ref{c13p}), which is dominated by the
lowest-order term, yields in the limit $a>>R=L/2$%
\[
\widetilde{C}(\alpha)\simeq\frac{2\pi}{\ln(32\alpha)},
\]
thereby confirming once again Eq.(\ref{c13n}).

A caution is required in the opposite, short-cylinder regime: some tables
approach a finite constant as $\alpha\rightarrow\infty$, which corresponds to
a disk-like (end-cap--dominated) conductor with finite thickness. In our
problem, the conductor is an open cylindrical shell (lateral surface only)
with zero wall thickness; in this model, the short-cylinder limit is ring-like
and singular, and a physical regularization requires an additional transverse
length scale (wall thickness or wire radius). Therefore, only the slender-body
entries ($x<<1$, $\alpha>>1$) should be used for direct quantitative
comparison with the present shell model.

\subsubsection{Capacitance Behavior Across Aspect Ratios}

Figure 4 summarizes the numerical behavior of the dimensionless capacitance
$\widetilde{C}(\alpha)$ over a broad range of aspect ratios $\alpha=a/L$. The
results provide a smooth quantitative connection between the slender-cylinder
limit $\alpha\rightarrow0$ ($L>>a$), where $\widetilde{C}(\alpha)$ grows
rapidly and is well described by the classical Maxwell logarithmic asymptote
in Eq. (\ref{c14}), and the short-cylinder limit $\alpha\rightarrow\infty$
($L<<a$), where the response becomes increasingly dominated by the rim regions
and the dependence on geometry is only logarithmic. The inset, showing
$1/\widetilde{C}(\alpha)$ versus $\ln\alpha$, makes this slow trend explicit
and confirms the expected large$-\alpha$ behavior of the ideal zero-thickness
shell, Eq. (\ref{c13n}).

\begin{figure}%
[p]
  \centering
  \includegraphics[width=0.86\linewidth]{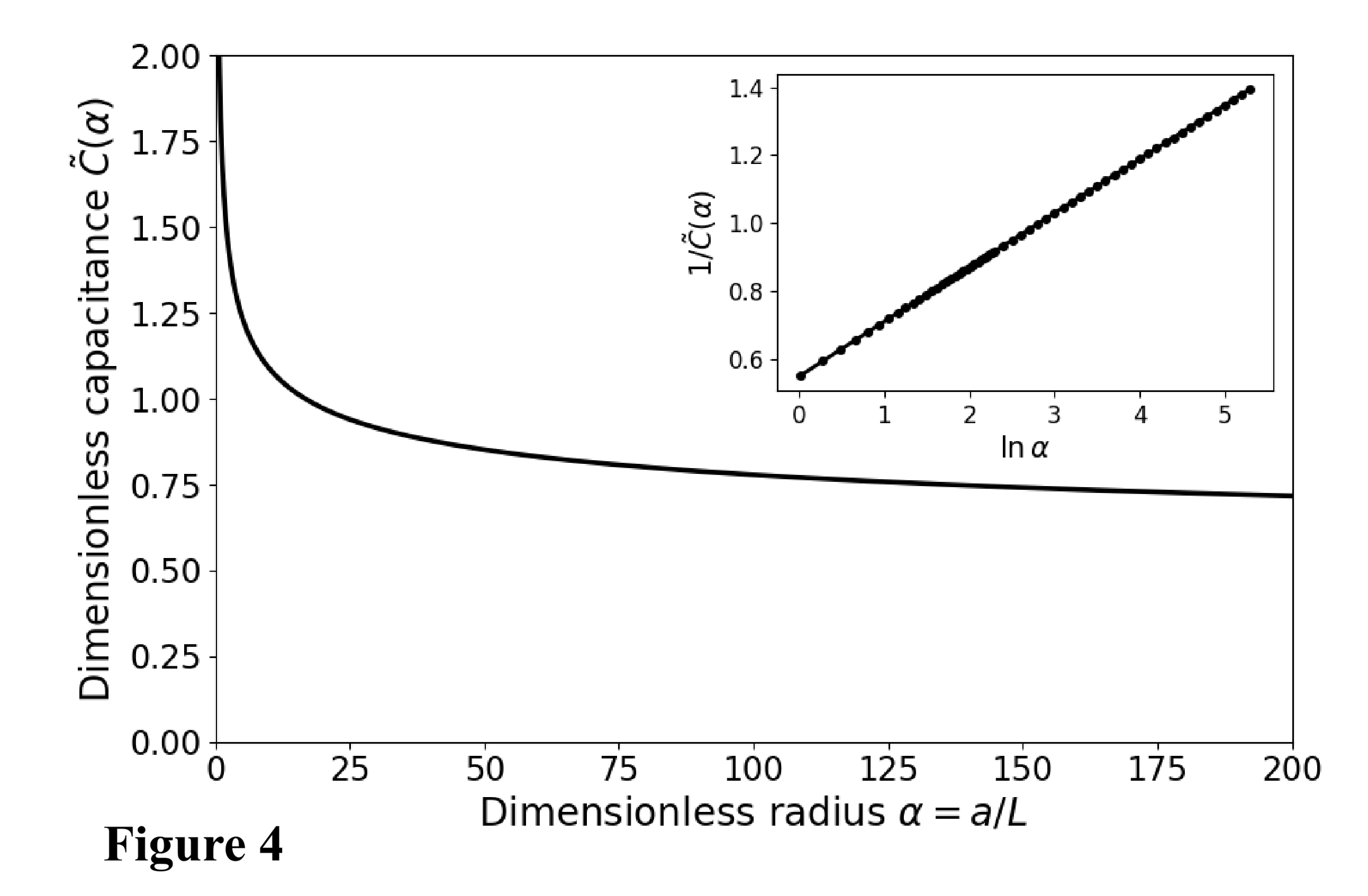}%

  \caption{Dimensionless total capacitance of an open conducting cylinder (lateral surface only) held at a uniform potential $V_0$, shown as a function of the aspect parameter $\alpha=a/L$. The plotted quantity is $C(\alpha)$ obtained from the numerically computed surface-charge distribution, Eq.~(\ref{c13}%
). The vertical range is restricted to $0\le C(\alpha)\le 2.0$ to emphasize the rapid growth as $\alpha\to 0$ (long-cylinder limit) while preserving detail at moderate $\alpha$. Inset: $1/C(\alpha)$ versus $\ln\alpha$, emphasizing the slow logarithmic trend in the short-cylinder regime $\alpha\gg 1$, see Eq.~(\ref{c13n}%
).}
  \label{fig:capacitance_thin}
\end{figure}

Beyond its pedagogical value, the present finite-length cylinder solution
provides quantitative benchmarks for end-dominated electrostatics that are
immediately useful in two applied settings. (i) \textbf{High-voltage hardware}
and \textbf{feedthrough engineering}: breakdown and surface flashover are
frequently initiated at geometric discontinuities (ends, triple-junction
regions, and abrupt curvature changes) where the local field is enhanced;
therefore, a controlled description of how the end singularity in $\sigma(z) $
and the geometric capacitance $C(\alpha)$ scale with aspect ratio offers a
compact way to parameterize and validate design choices in high-voltage
systems (including electrode shaping and insulating feedthrough
geometries)\cite{ap1,ap2,ap3,ap4,ap5}. (ii) \textbf{Capacitance metrology} and
\textbf{guarded electrode geometries}: guard-ring and re-entrant designs are
employed precisely to suppress or correct fringing fields, and the residual
fringing contribution is often the limiting systematic when cylindrical
electrodes are used; thus, $C(\alpha)$ and $\sigma(z)$ supply
geometry-resolved inputs for validating FEM/BEM corrections and for
constructing/calibrating guarded standards and calculable-capacitor
uncertainty budgets\cite{ap6,ap7,ap8,ap9}.

More broadly, the asymptotic form (\ref{c14}) quantifies the familiar ``nearly
linear'' scaling of capacitance with length in slender conductors while making
explicit the slow logarithmic correction that can matter at realistic aspect
ratios. This correction is often invoked qualitatively in engineering
approximations, but the present treatment offers a controlled bridge between
the slender regime and the short, edge-dominated regime, providing a benchmark
curve $C(\alpha)$ that can be used to validate reduced models,
boundary-element solvers, or simplified circuit representations of finite electrodes.

The capacitance provides a compact global characterization of the
electrostatic response of the finite cylindrical conductor. Because it
integrates the full surface-charge distribution, it is sensitive both to the
bulk behavior and to the rim singularities discussed in the preceding
asymptotic analysis. As such, the dependence of $C(\alpha)$ on the aspect
ratio offers a stringent diagnostic of the numerical solution and a natural
way to connect the different physical regimes of the problem.

\section{Spectral Formulation}

Within the spectral approach, the electrostatic potential in the vacuum region
outside the conducting cylindrical shell is represented by a Hankel
(Fourier--Bessel) superposition of axisymmetric Laplace modes, namely%

\begin{equation}
\Phi(\rho,z)=\int\limits_{0}^{\infty}A(k)J_{0}(k\rho)e^{-k\left|  z\right|
}dk, \label{c15}%
\end{equation}
where $A(k)$ is an a priori unknown spectral coefficient encoding the finite
geometry of the conductor. This representation follows directly from
separation of variables for the axisymmetric Laplace equation in cylindrical
coordinates: for each transverse wavenumber $k$, $J_{0}(k\rho)$ provides the
regular radial eigenfunction at $\rho=0$, while the axial dependence
$e^{-k\left|  z\right|  }$ selects the decaying solution away from the plane
$z=0$ and guarantees regularity at infinity. The absolute value $\left|
z\right|  $ enforces the even symmetry in $z$ appropriate to the present
geometry, and the continuous integral over $k$ reflects the unbounded radial
domain. Closely related Hankel representations have been used in classical
dual--integral--equation treatments of the hollow finite
cylinder\cite{c2a,c2b,c2c,c3,c4}.

Imposing the boundary condition on the surface of the cylindrical shell,
$\Phi(a,z)=V_{0}$ for $\left|  z\right|  <L/2$, Eq. (\ref{c15}) reduces to the
integral equation%
\begin{equation}
\int\limits_{0}^{\infty}A(k)J_{0}(ka)e^{-k\left|  z\right|  }dk=V_{0}\text{,
}\left|  z\right|  <L/2. \label{c16}%
\end{equation}
Equation (\ref{c16}) is a \textit{Fredholm integral equation} of the first
kind for the spectral coefficient $A(k)$. Because it is valid only on a finite
interval of the axial coordinate, its inversion is nontrivial and, in the
classical literature, leads to a coupled system of dual integral equations
when supplemented with the regularity conditions outside the
conductor\cite{c2a,c2b,c2c,c3,c4}.

Rather than determining $A(k)$ directly, it is advantageous to relate it to
the physically meaningful surface-charge density $\sigma(z)$. The latter is
obtained from the normal derivative of the potential at the surface,%
\begin{equation}
\sigma(z)=-\varepsilon_{0}\left.  \frac{\partial\Phi}{\partial\rho}\right|
_{\rho=a}. \label{c17}%
\end{equation}
Using Eq. (\ref{c15}) and the indentity $J_{0}^{\prime}(k\rho)=-kJ_{1}(k\rho
)$, one finds%
\begin{equation}
\sigma(z)=\varepsilon_{0}\int\limits_{0}^{\infty}kA(k)J_{1}(ka)e^{-k\left|
z\right|  }dk, \label{c18}%
\end{equation}
where $J_{1}(x)$ is the Bessel function of first order. Equation (\ref{c18})
shows that the spectral coefficient $A(k)$ plays the role of a Laplace--Hankel
transform of the surface-charge density. Formally inverting this relation, one
may write%
\begin{equation}
A(k)=\frac{1}{\varepsilon_{0}J_{1}(ka)}\int\limits_{-L/2}^{L/2}\sigma
(z)e^{-k\left|  z\right|  }dk, \label{c19}%
\end{equation}
which expresses the spectral coefficient explicitly in terms of the unknown
charge density $\sigma(z)$. This step is standard in spirit (it mirrors the
classical dual-integral-equation constructions\cite{c3,c4}), but here we use
it to eliminate $A(k)$ entirely in favor of $\sigma(z)$.

Substituting Eq. (\ref{c19}) into the boundary condition (\ref{c16}) and
interchanging the order of integration, we obtain a closed integral equation
for $\sigma(z)$:%

\begin{equation}
\int\limits_{-L/2}^{L/2}\sigma(z^{\prime})\left[  \int\limits_{0}^{\infty
}\frac{J_{0}(ka)}{J_{1}(ka)}e^{-k\left|  z-z^{\prime}\right|  }dk\right]
e^{-k\left|  z\right|  }dk=\varepsilon_{0}V_{0}. \label{c20}%
\end{equation}
The problem is thus reduced, within the spectral framework, to a single
real-space integral equation for $\sigma(z)$, with a kernel given by the
inverse spectral integral in brackets. In the classical dual-integral-equation
literature\cite{c3,c4}, the analysis typically proceeds by manipulating Eq.
(\ref{c16})$-$(\ref{c18}) directly in Bessel space. Our aim here is different:
we use the spectral representation only as a device to recover, in a
systematic way, the same real-space operator that was obtained in Sec. 2 by
direct ring-to-ring integration.

The kernel appearing in Eq. (\ref{c20}) can be evaluated explicitly using
standard integral identities involving Bessel functions. In particular, one
finds%
\begin{equation}
\int\limits_{0}^{\infty}\frac{J_{0}(ka)}{J_{1}(ka)}e^{-k\left|  z-z^{\prime
}\right|  }dk=\frac{1}{a}\frac{K\left(  \frac{4a^{2}}{4a^{2}+\left(
z-z^{\prime}\right)  ^{2}}\right)  }{\sqrt{4a^{2}+\left(  z-z^{\prime}\right)
^{2}}}, \label{c21}%
\end{equation}
where $K(m)$ denotes the complete elliptic integral of the first kind, and the
identity may be traced to standard tables of Bessel integrals\cite{tab}.
Substitution of Eq. (\ref{c21}) into Eq. (\ref{c20}) yields precisely the
real-space integral equation (\ref{c5}) derived earlier from ring-to-ring
interactions. In other words, Eqs. (\ref{c5}) and (\ref{c16}) provide two
mathematically equivalent formulations of the same electrostatic
boundary-value problem: the former in terms of an elliptic kernel in physical
space, the latter in terms of a spectral amplitude $A(k)$ and dual integral equations.

The spectral formulation, based on Bessel-function expansions, emphasizes the
modal structure of the solution and naturally leads to dual integral equations
as in Refs.\cite{c2a,c2b,c2c,c3,c4}. The real-space formulation, obtained here
by explicitly eliminating the spectral coefficient and evaluating the Bessel
integral, makes the nonlocal interaction along the cylindrical surface fully
explicit through a kernel involving elliptic integrals. The equivalence
between these approaches follows directly from the inversion of the
Hankel--Laplace transform and demonstrates that the elliptic-integral kernel
is nothing but the real-space representation of the underlying spectral
operator. In both formulations, the finite length of the cylinder manifests
itself through the singular behavior of the kernel, leading to an integrable
divergence of the surface-charge density near the edges $z=\pm L/2$. This
behavior is independent of the chosen representation and reflects the
universal nature of edge effects in electrostatics. While both viewpoints
appear separately in the literature, the present derivation provides an
explicit elimination step that yields the elliptic kernel in closed form and
is subsequently used in Sec. 3 to design a rim-aware discretization with
controlled extrapolation.

\section{Conclusion}

In this work, we investigated the electrostatics of a thin conducting
cylindrical shell of finite length maintained at a uniform potential. The
problem was first formulated directly in real space, leading to an integral
equation for the surface-charge density whose kernel is expressed in terms of
complete elliptic integrals. This formulation provides a physically
transparent description of the interaction between different axial elements of
the conductor and proves to be particularly convenient for numerical
implementation. The resulting charge distribution correctly captures the
strong charge accumulation near the cylinder ends, as well as its dependence
on the geometric aspect ratio $L/a$.

Subsequently, we revisited the same problem using a spectral representation of
the electrostatic potential in terms of Bessel functions. By explicitly
eliminating the spectral coefficient, we demonstrated the exact equivalence
between the spectral formulation and the real-space integral equation with an
elliptic kernel. This equivalence clarifies the relationship between two
commonly employed but often separately presented approaches and shows that
they are simply different representations of the same underlying electrostatic problem.

The numerical results obtained from the integral-equation approach allow for
the accurate determination of the total induced charge and, consequently, of
the capacitance of the finite conducting cylinder. Such configurations are of
practical interest in a variety of applications, including the modeling of
electrodes in vacuum devices, nanoscale conductive structures, cylindrical
sensors, and components of capacitive energy-storage systems. Moreover, the
methodology developed here can be readily extended to more general boundary
conditions, non-uniform potentials, or coupled multi-conductor geometries.

Overall, the present analysis provides a unified and consistent framework for
the electrostatics of finite cylindrical conductors, combining physical
transparency, mathematical rigor, and numerical efficiency. We expect that the
equivalence established between the elliptic-kernel formulation and the
spectral Bessel representation will be useful in both analytical studies and
computational implementations of related electrostatic problems.

In addition to these possible extensions, the present results do more than
solve a classical boundary-value problem: they supply benchmark-quality data
for both the axial surface-charge profiles $\sigma(z/L)$ and the dimensionless
capacitance $\widetilde{C}(\alpha)$ over a wide range of aspect ratios.
Because the rim singularity is built into the numerical ansatz and the
convergence is quantified via finite-size scaling, these results provide a
stringent test for high-accuracy boundary-element, finite-element, and
method-of-moments solvers in electrostatics. In particular, any numerical
scheme that aims to resolve edge fields on finite cylindrical conductors can
be validated against the reference profiles reported here, both at the level
of global observables (capacitance) and local behavior (near-edge charge
build-up). In this sense, the finite conducting cylinder, equipped with the
elliptic-kernel integral formulation and Chebyshev-weighted discretization,
serves as a reusable benchmark geometry for future developments in
computational electrostatics.

\section*{Declaration of competing interest}

The author declares that there are no known competing financial interests or
personal relationships that could have appeared to influence the work reported
in this paper.

\section*{Acknowledgements}

The author acknowledges the use of an AI-based language model (ChatGPT,
OpenAI) to assist in improving the English language and clarity of some parts
of this manuscript. All scientific content, results, and interpretations are
the sole responsibility of the author.

\end{document}